\documentclass[floatfix,twocolumn,aps,prd,superscriptaddress,longbibliography]{revtex4-1}
\usepackage{graphicx}
\usepackage{dcolumn}
\usepackage{bm}
\usepackage{textcomp}
\usepackage[intlimits]{amsmath}
\usepackage{esint}
\usepackage{braket}
\usepackage{color}

\usepackage{soul} 

\soulregister\ref{7}
\soulregister\eqref{7}
\soulregister\cite{7}
\soulregister\onlinecite{7}

\usepackage{xspace}
\newcommand{\Lim}{Lima\c{c}on\xspace}

\begin{document}
	
\title{Spatio-temporal lasing dynamics in a Lima\c{c}on-shaped microcavity}

\author{Kyungduk Kim}
\affiliation{Department of Applied Physics, Yale University, New Haven, Connecticut 06520, USA}%
\author{Stefan Bittner}
\affiliation{Chair in Photonics, CentraleSup\'elec, LMOPS, 2 rue Edouard Belin, Metz 57070, France}%
\affiliation{Universit\'e de Lorraine, CentraleSup\'elec, LMOPS, 2 rue Edouard Belin, Metz 57070, France}%
\author{Yuhao Jin}
\affiliation{Center for OptoElectronics and Biophotonics, School of Electrical and Electronic Engineering, School of Physical and Mathematical Science, and Photonics Institute, Nanyang Technological University, 639798, Singapore}%
\author{Yongquan Zeng}
\affiliation{Center for OptoElectronics and Biophotonics, School of Electrical and Electronic Engineering, School of Physical and Mathematical Science, and Photonics Institute, Nanyang Technological University, 639798, Singapore}%
\author{Qi Jie Wang}
\affiliation{Center for OptoElectronics and Biophotonics, School of Electrical and Electronic Engineering, School of Physical and Mathematical Science, and Photonics Institute, Nanyang Technological University, 639798, Singapore}%
\author{Hui Cao}
\email{hui.cao@yale.edu}
\affiliation{Department of Applied Physics, Yale University, New Haven, Connecticut 06520, USA}%

\begin{abstract}
	\Lim-shaped microdisk lasers are promising on-chip light sources with low lasing threshold and unidirectional output. We conduct an experimental study on the lasing dynamics of \Lim-shaped semiconductor microcavities. The edge emission exhibits intensity fluctuations over a wide range of spatial and temporal scales. They result from multiple dynamic processes with different origins and occurring on different spatio-temporal scales. The dominant process is an alternate oscillation between two output beams with a period as short as a few nanoseconds.
\end{abstract}

\maketitle 

	
	Dielectric microdisk cavities are widely exploited for modern optoelectronic devices due to their strong confinement of light by total internal reflection. Long photon lifetimes in circular disks lead to high quality (Q) factors and low lasing thresholds, but the isotropic emission results in poor collection efficiency \cite{cao2015dielectric}. Directional output and improved light extraction can be achieved with deformed disks~\cite{schwefel2004progress, harayama2011two, cao2015dielectric}. Among the microdisk geometries that have been explored, the \Lim cavity has the advantage of combining unidirectional emission and high-Q factor~\cite{wiersig2008combining}. There have been many experimental and theoretical studies of the far-field emission pattern~\cite{wang2009deformed, shinohara2009ray, song2009chaotic, yan2009directional, yi2009lasing, song2010directional, albert2012directional, redding2012local, kreismann2017three} and the mechanism for the formation of high-Q resonances in the \Lim-shaped microcavities in which the classical ray dynamics is predominantly chaotic~\cite{shim2011whispering, shim2013adiabatic, kraft2014perturbative}. However, most of these studies are focused on the static properties of cavity resonances and steady-state lasing emission. 
	
	Despite the potential of the \Lim-shaped microdisk laser as a highly efficient on-chip light source, its temporal dynamics has not yet been studied. For edge-emitting semiconductor lasers, the resonator geometry can have a profound impact on the lasing dynamics~\cite{choi2008alternate, shinohara2014anticorrelated}. Carefully designed cavity shapes for semiconductor quantum-well lasers can suppress the intrinsic spatio-temporal instabilities~\cite{bittner2018suppressing,kim2021massively,kim2022sensitive} and control the nonlinear light-matter interactions~\cite{ma2022chaotic}. 
	
	In this work, we experimentally investigate the spatio-temporal dynamics of \Lim-shaped microdisk lasers. We observe highly multimode lasing under electrical pumping of a GaAs quantum-well. The interactions of many lasing modes and the gain material result in multiple dynamic processes such as anti-phase oscillations, temporal instabilities, and spatio-temporal mode beating. These processes have distinct spatial and temporal scales and coexist in a single cavity. Among them, the dominant process in the \Lim laser is the anti-correlated oscillation of two output beams, with oscillation periods as short as a few nanoseconds. For comparison, the lasing dynamics in an elliptic microcavity with integrable ray dynamics is dominated by filament-like instabilities.

\begin{figure}[t!]
		\centering\includegraphics[width=0.9\linewidth]{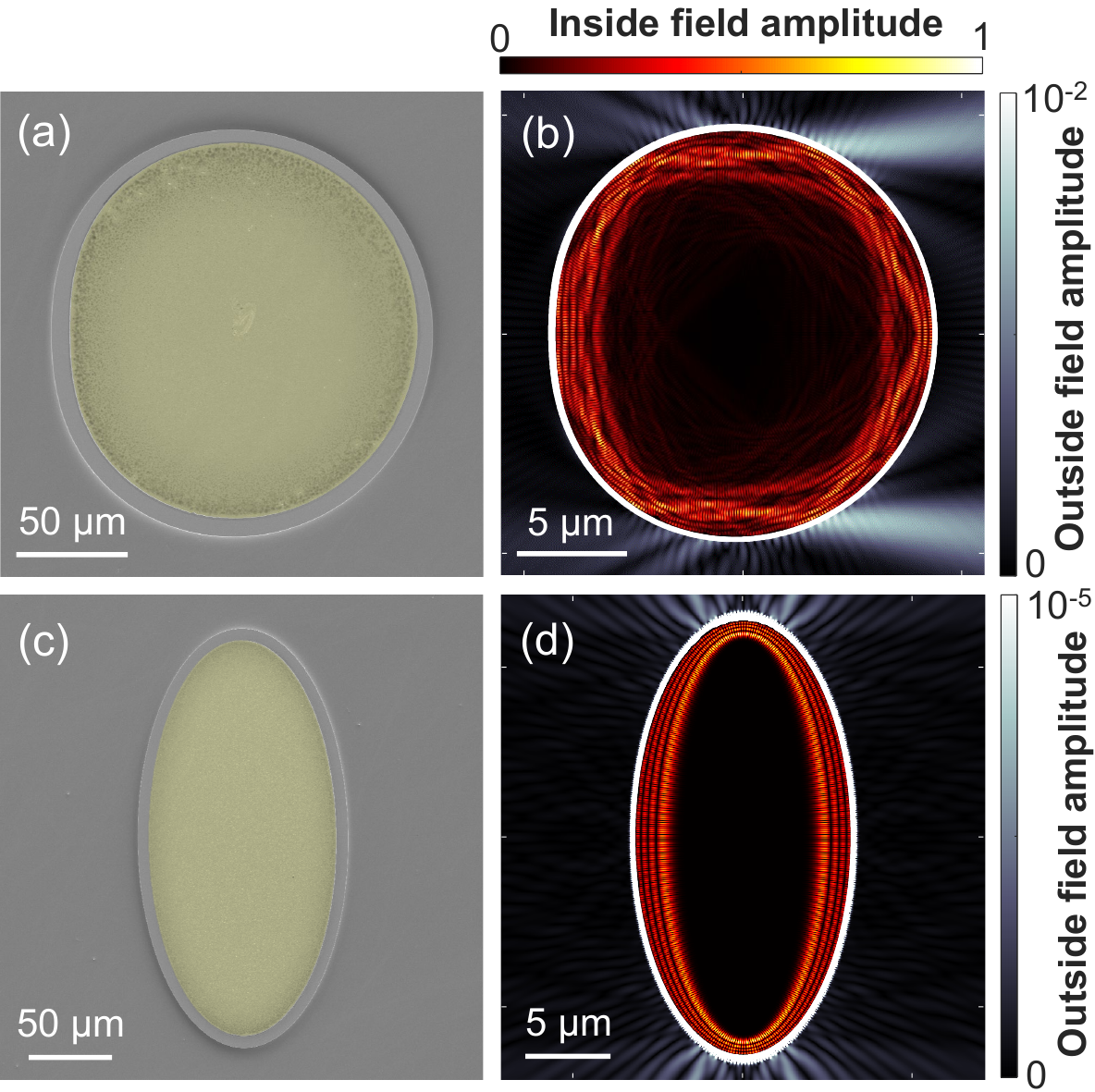}
		\caption{Semiconductor microdisk lasers. (a,c) SEM images of \Lim-shaped (a) and elliptic (c) microdisks. The yellow area represents the top metal contact for the current injection. (b,d) Numerically calculated high-Q modes in both cavities. They concentrate near the cavity boundaries, and total internal reflection leads to strong optical confinement. The mode in the \Lim cavity (b) has two major output beams in the same direction. In the elliptic cavity (d), evanescent tunneling is strongest at the points with the smallest radius of curvature, producing four output beams in two directions.}
		\label{fig1}
	\end{figure}

	The \Lim boundary is given in polar coordinates by $r(\theta) = R(1+\varepsilon\cos{\theta})$, where $R$ is the radius and $\varepsilon$ is the deformation parameter. We choose $\varepsilon = 0.42$ to obtain high-Q modes in a GaAs microdisk with unidirectional output~\cite{wiersig2008combining, wang2009deformed, shinohara2009ray}. Figure~\ref{fig1}(a) is the scanning electron microscope (SEM) image of a \Lim-shaped disk with $R$ = 86 $\mu$m. It is fabricated with a commercial GaAs/AlGaAs quantum-well wafer (Q-Photonics QEWLD-808) by photolithography and inductively coupled plasma etching \cite{kim2022sensitive}. Top and bottom metal contacts are deposited for electric current injection (See Supplement 1). 
	
	In spite of chaotic ray dynamics in the \Lim cavity, high-Q modes can be formed by scarring on unstable periodic orbits and dynamic localization~\cite{wiersig2008combining,song2009chaotic,shim2011whispering}. Figure~\ref{fig1}(b) shows the spatial intensity distribution of a high-Q mode with transverse-electric (TE) polarization, obtained by numerical simulation of a \Lim disk with $R$ = 8.6 $\mu$m. Similar to the whispering gallery (WG) modes in a circular cavity, the high-Q resonance stays close to the \Lim cavity boundary but has a more complex structure due to the chaotic ray dynamics. In phase space, the high-Q modes are concentrated in the region confined by total internal reflection, leading to long lifetimes. Eventually, light will reach the leaky region with an incident angle on the boundary below the critical angle. As light leaves the cavity via refraction, the escape route is dictated by the unstable manifolds of the chaotic saddle in phase space, producing two output beams in a single direction, as shown in Fig.~\ref{fig1}(b).
	
	To investigate the effect of ray dynamics on lasing dynamics, we fabricate elliptic microdisks with integrable ray dynamics on the same wafer. Figure~\ref{fig1}(c) shows an elliptic disk with the major (minor) axis length $b$ = 254 $\mu$m ($a$ = $b/2$ = 127 $\mu$m). Its area is identical to that of the \Lim disk. The high-Q resonances in an elliptic cavity are whispering-gallery modes formed by total internal reflection~\cite{nockel1996directional, kim2004highly, schwefel2004dramatic}. Figure~\ref{fig1}(d) shows the spatial distribution of a high-Q resonance obtained by numerical simulation of an elliptic disk with $a$ = 12.7 $\mu$m and $b$ = 25.4 $\mu$m. It is a WG mode with well-defined radial and azimuthal quantum numbers. Light leaks via tunneling at the curved boundary. The tunneling rate is highest at the smallest radius of curvature. Thus, the emission is strongest at the two vertices [Fig.~\ref{fig1}(d)], leading to four output beams in two opposite directions.
	
\begin{figure}[b!]
		\centering\includegraphics[width=0.95\linewidth]{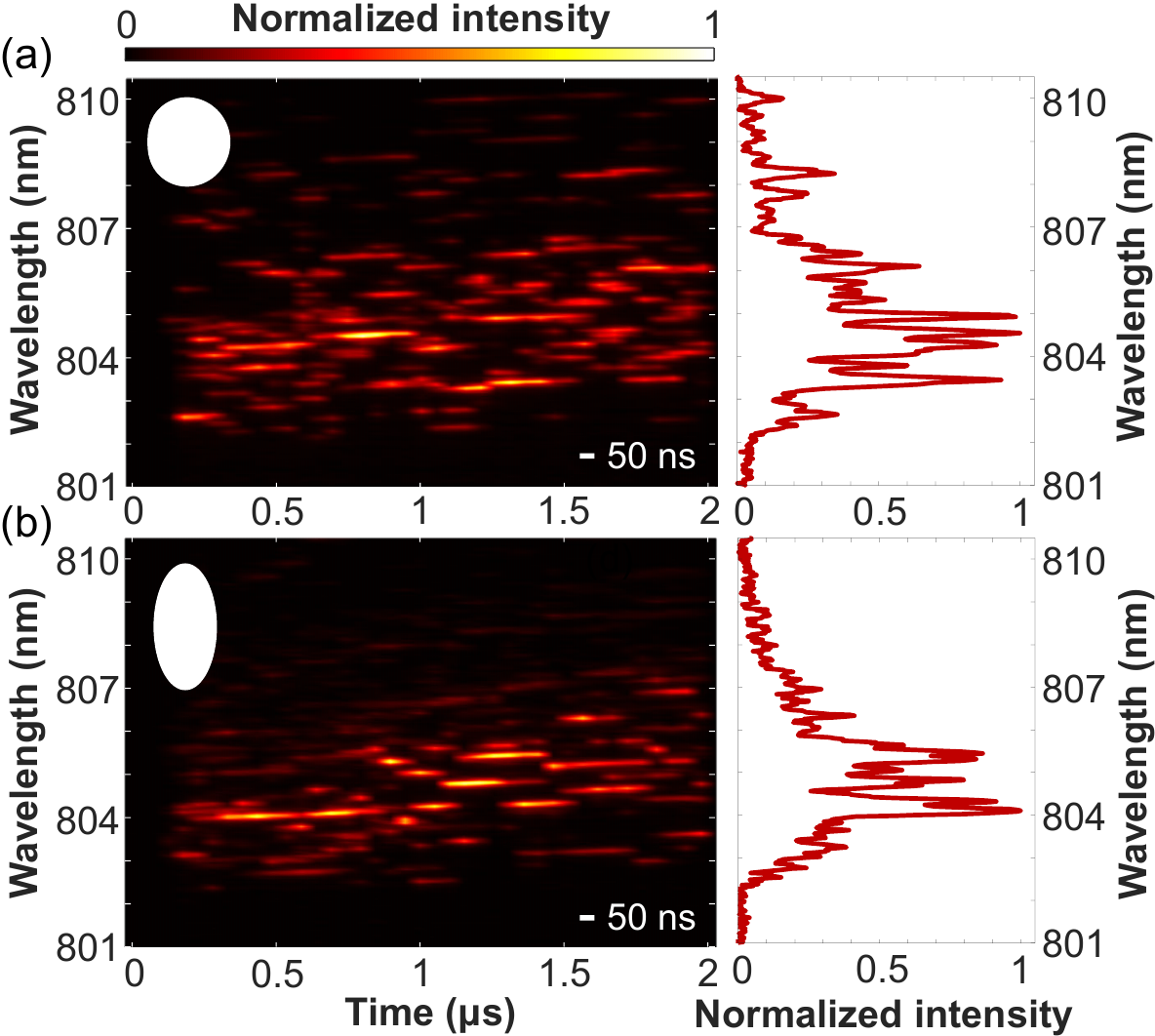}
		\caption{Highly multimode lasing in \Lim and elliptic microdisks. Left panels: time-resolved emission spectra reveal multimode lasing in both \Lim (a) and elliptic (b) cavities at any time during 2~$\mu$s long pulses. The spectrum redshifts and lasing peaks change over time due to heating. 
		Right panels: time-integrated spectra exhibiting many lasing peaks.}
		\label{fig2}
	\end{figure}

	We achieve lasing of both \Lim and elliptic microdisks at room temperature by electrical pumping. To reduce heating, the electric current pulses are only 2~$\mu$s long (See Supplement 1). The lasing threshold current is 80~mA for both \Lim and elliptic cavities. Below we present data taken at a pump current of 500~mA. Figure~\ref{fig2} (right panels) shows the time-integrated emission spectra centered at wavelength 805~nm. For both \Lim and elliptic microdisk lasers, the emission spectra contain many peaks, indicating highly multimode lasing. The left panels of Fig.~\ref{fig2} are time-gated emission spectra measured by an intensified CCD camera. Due to the heating effect, the lasing spectra redshift, and some existing peaks disappear while new ones appear. Spectral decorrelation occurs on the time scale of 30-40 ns (See Supplement 1). The lasing spectra contain multiple peaks at any moment, confirming simultaneous lasing of multiple modes in both \Lim and elliptic cavities.
	
	Next, we study the spatio-temporal dynamics of laser emission using a streak camera (See Supplement 1). Figure~\ref{fig3}(a) shows the time-resolved near-field intensity profile of a \Lim-shaped microdisk laser. It features two dominant output beams, as expected from the output of high-Q modes [Fig.~\ref{fig1}(b)]. Both beam intensities fluctuate in time on multiple scales. The time traces (inset) reveal rapid oscillations on top of a slow-varying background. Within the first 5 ns time window, the beams at the position $x > 0$ is stronger than the other beam at $x < 0$, but in the next 5 ns, the $x < 0$ beam becomes stronger than the $x > 0$ beam. 
	
	\begin{figure}[t!]
		\centering\includegraphics[width=\linewidth]{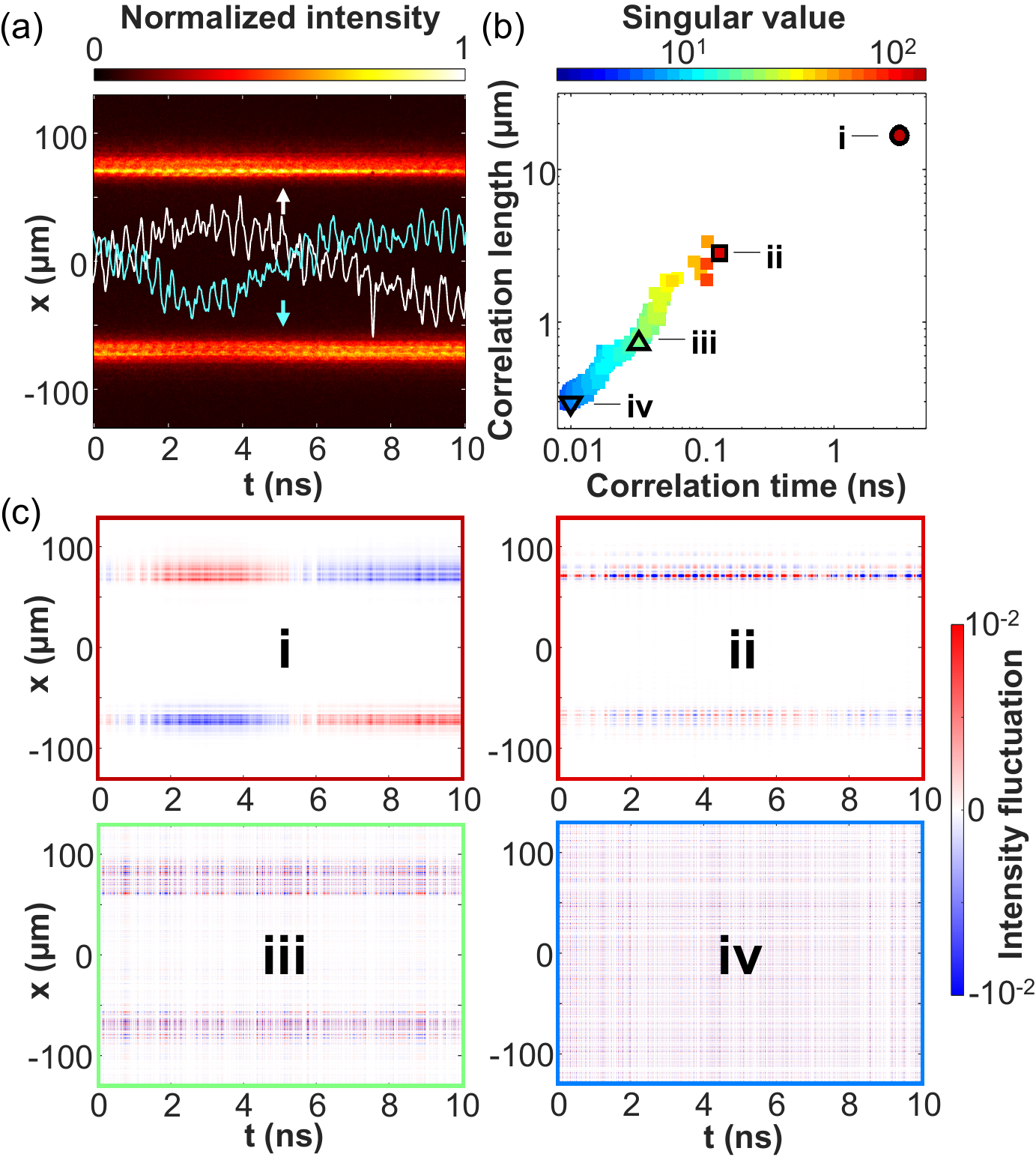}
		\caption{Spatio-temporal dynamics of lasing in a \Lim-shaped microcavity. (a) Spatio-temporal distribution of near-field emission intensity measured by a streak camera with a temporal resolution of about 30 ps. Solid lines represent the intensity traces of the two output beams. (b) Correlation length vs. correlation time of individual singular vectors constituting the emission pattern in (a). The magnitude of their singular values $s_{\alpha}$ is indicated by color. (c) Four spatio-temporal singular vectors with distinct spatio-temporal scales marked by \textbf{i}-\textbf{iv} in (b). \textbf{i}: anti-correlated oscillations of two output beams ($\alpha$ = 1). \textbf{ii}:  spatial filaments and irregular pulsations ($\alpha$ = 2). \textbf{iii}:  spatio-temporal beating of lasing modes ($\alpha$ = 45). \textbf{iv}:  photo-detection noise ($\alpha$ = 300).}
		\label{fig3}
	\end{figure}
	
	To separate dynamic processes of different spatio-temporal scales, we perform a singular value decomposition (SVD) (equivalent to a Karhunen-Loeve decomposition~\cite{hess1994spatio}) of the measured intensity fluctuations. Consider a streak image $I(x,t)$, after normalization $\langle I(x,t) \rangle_{x,t} = 1$, we compute the time-averaged intensity $\langle I(x,t) \rangle_t$ at each spatial location $x$, then obtain the spatially-resolved intensity fluctuation in time $\delta I(x,t) = I(x,t) - \langle I(x,t) \rangle_t$. The SVD of $\delta I(x,t)$ yields the decomposition $\delta I(x,t) = \sum_{\alpha} s_{\alpha} u_{\alpha}(x) v_{\alpha}(t)$, where $s_{\alpha}$ is the singular value, and $u_{\alpha}(x)$ and $v_{\alpha}(t)$ denote spatial and temporal singular vectors, respectively. The singular value represents the contribution of the corresponding singular vector to the intensity fluctuation. The lower the index $\alpha$, the larger the singular value $s_{\alpha}$, and the more dominant the singular vector $u_{\alpha}(x) v_{\alpha}(t)$. Each singular vector has its own characteristic scales in space and time. The spatial scale is given by the correlation length $l_{\alpha}$, which is defined as the full-width-at-half-maximum (FWHM) of the spatial correlation function for $u_{\alpha}(x)$~\cite{kim2022sensitive}. Correspondingly, the temporal scale is the correlation time $\tau_{\alpha}$ given by the FWHM of the temporal correlation function for $v_{\alpha}(t)$. 
	
	The SVD effectively separates a variety of spatial and temporal scales of the laser intensity fluctuations. As shown in Fig.~\ref{fig3}(b), the singular vectors feature a wide range of correlation times $\tau_{\alpha}$ and lengths $l_{\alpha}$. The most dominant singular vector is denoted as ($\mathbf{i}$). It has $\tau_{\mathbf{i}}$ = 3.1 ns and $l_{\mathbf{i}}$ = 17 $\mu$m, which are much larger than the spatial and temporal scales of other singular vectors. The corresponding singular vector $u_{\mathbf{i}}(x)v_{\mathbf{i}}(t)$ exhibits anti-correlated intensity oscillations of two output beams.
	
	The next dominant singular vector ($\mathbf{ii}$) displays more rapid fluctuations in space and time. It has $\tau_{\mathbf{ii}}$ = 0.13 ns and $l_{\mathbf{ii}}$ = 2.8 $\mu$m. The two major output beams break into finer streaks of a few microns, resembling filaments in an edge-emitting semiconductor laser \cite{kim2022sensitive}. The sub-nanosecond fluctuation time is very similar to that of irregular pulsations induced by filamentation in a GaAs quantum-well laser~\cite{fischer1996complex,marciante1998spatio,bittner2018suppressing}. Such spatio-temporal dynamics is therefore attributed to lasing instabilities caused by nonlinear light-matter interaction.
	
	The majority of singular vectors have even shorter spatial and temporal scales. Since their singular values are small, individual contributions to intensity fluctuations are weak. We show two examples, ($\mathbf{iii}$) and ($\mathbf{iv}$) in Fig.~\ref{fig3}(c). The spatial profile of the singular vector ($\mathbf{iii}$) resembles the emission pattern of the \Lim with two output beams, thus it originates from lasing emission. Its $\tau_{\mathbf{iii}}$ and $l_{\mathbf{iii}}$ are 33 ps and 0.72 $\mu$m, close to the temporal and spatial resolution of our photo-detection system. According to our previous studies~\cite{kim2021massively}, the spatio-temporal profile of ($\mathbf{iii}$) results from the interference of multiple lasing modes in space and time. Lastly, the singular vector ($\mathbf{iv}$) spreads uniformly over the entire spatio-temporal domain of detection and has the correlation scales of a single camera pixel. Hence, it corresponds to photo-detection noise \cite{kim2022sensitive}.

	
	Figure~\ref{fig3} illustrates the SVD is a powerful tool for analyzing spatio-temporal processes of different origins. However, the singular vectors in Fig.~\ref{fig3}(b) are extracted from a single 10~ns long streak image. A lot more sampling is needed for a statistical analysis of the overall lasing dynamics. To this end, we perform the SVD of many streak images taken with multiple devices. Within a 2~$\mu$s long pump pulse, 161 consecutive 10~ns long streak images are taken from 0.3 to 1.9 $\mu$s. The measurements are repeated with five cavities of the same geometry fabricated on the same wafer. We perform the SVD of $161\times5=805$ streak images in total and combine the results.
	
\begin{figure}[t!]
		\centering\includegraphics[width=\linewidth]{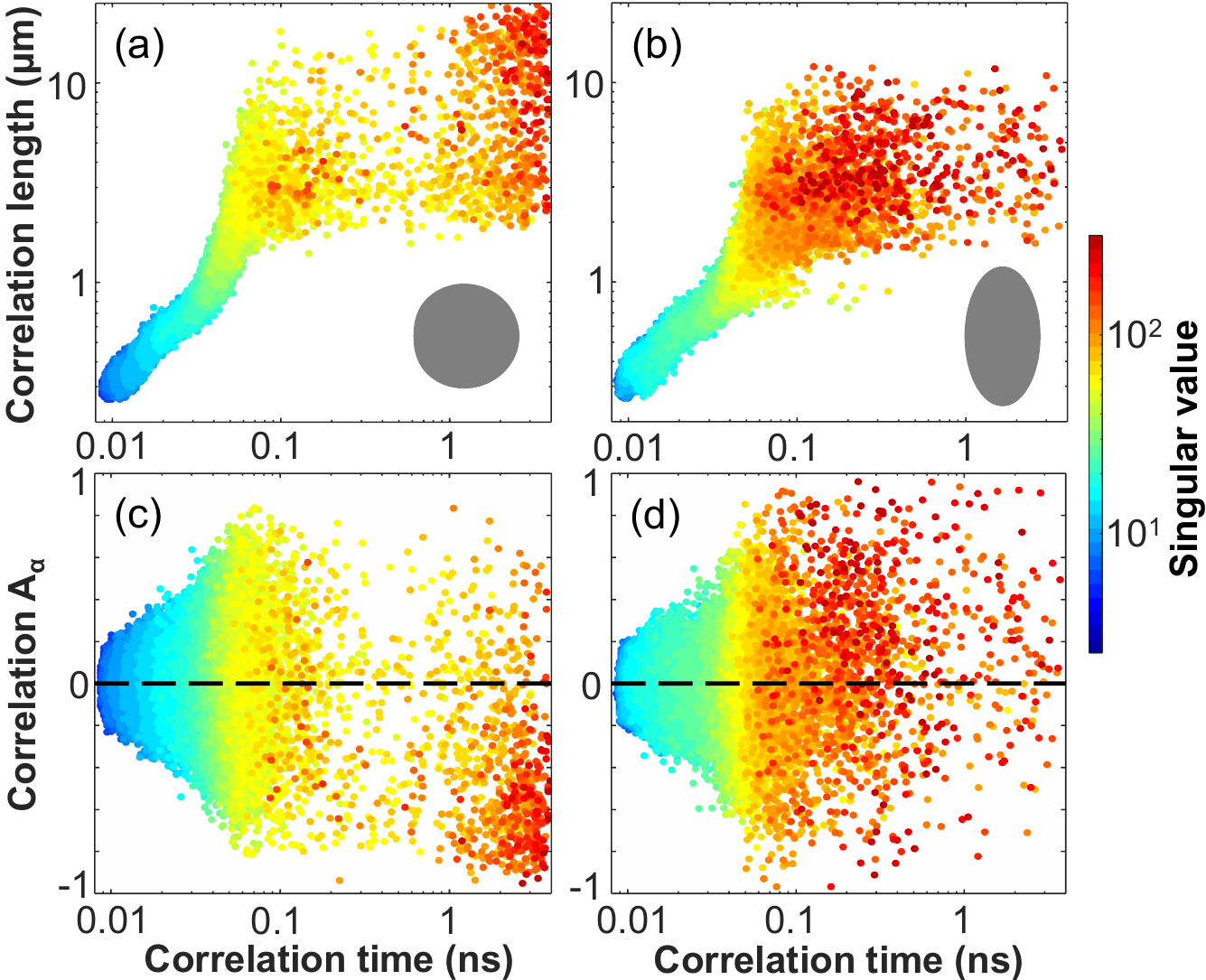}
		\caption{Statistical analysis of spatio-temporal scales of lasing intensity fluctuations. (a, b) Spatial and temporal correlation widths of all singular vectors for emission intensity fluctuations of five \Lim-shaped (a) and elliptic (b) microdisk lasers. The vectors with large singular values (in red color) concentrate at a few nanoseconds for \Lim cavities, and around 0.1 ns for elliptic cavities. (c, d) Correlation $A_\alpha$ of temporal fluctuations of emission intensities between $x>0$ and $x<0$ of individual singular vectors. In \Lim-shaped microcavity lasers, the most dominant dynamics feature anti-correlated intensity fluctuations around a few nanoseconds, which are absent in elliptic lasers.}
		\label{fig4}
	\end{figure}

	Figure~\ref{fig4}(a) shows correlation times $\tau_{\alpha}$ and correlation lengths $l_{\alpha}$ of all singular vectors of five \Lim-shaped microdisk lasers. The color denotes the magnitude of singular value $s_\alpha$. The singular vectors can be classified into four categories, as represented by (\textbf{i}) - (\textbf{iv}) in Fig.~\ref{fig3}(c). The singular vectors with the largest $s_\alpha$ have $\tau_{\alpha}$ of a few nanoseconds and $l_{\alpha}$ from a few to tens of micrometers. Their spatio-temporal profiles resemble the singular vector ($\mathbf{i}$) in Fig.~\ref{fig3}. In addition, a small number of singular vectors with sub-nanosecond $\tau_{\alpha}$ have large singular values, and their profiles are similar to the singular vector ($\mathbf{ii}$) in Fig.~\ref{fig3}.  
	
	For comparison, we repeat the spatio-temporal measurement and SVD analysis with five elliptic microdisk lasers [Fig.~\ref{fig1}(c)]. In contrast to the case of \Lim-shaped microlasers, the strongest singular vectors belong to category (ii) with $\tau_{\alpha}$ of sub-nanoseconds and $l_{\alpha}$ of a few micrometers as shown in Fig.~\ref{fig4}(b). They result from filament-like spatio-temporal instabilities, which dominate the lasing dynamics in elliptic cavities. Comparison to \Lim-shaped lasers reveals an order-of-magnitude difference in the most-significant spatial and temporal scales of emission intensity fluctuations, illustrating the profound impact of cavity geometry on lasing dynamics. 
	
	To further characterize the dominant dynamic process in \Lim-shaped microdisk lasers, we investigate the correlation of intensity oscillations between the two output beams. For individual spatio-temporal singular vectors $u_{\alpha}(x) \, v_{\alpha}(t)$, temporal variations at different locations are either in phase or out of phase, leading to positive or negative correlations. We introduce the correlation for the $\alpha$-th singular vector as 
	\begin{equation}
	\label{pearson_AntiPhase} 
	A_{\alpha} = \frac{\langle [ u_\alpha(x) - \langle u_\alpha(x) \rangle ] [ u_\alpha(-x) - \langle u_\alpha(-x) \rangle ] \rangle}{\sqrt{  \langle  [ u_\alpha(x) - \langle u_\alpha(x) \rangle ]^2  \rangle \langle  [ u_\alpha(-x) - \langle u_\alpha(-x) \rangle ]^2  \rangle  }},
	\end{equation}
	where $\langle\cdots\rangle$ denotes the spatial average over $x>0$ only. Figure~\ref{fig4}(c) shows $A_{\alpha}$ for singular vectors of \Lim microdisk lasers. The most significant singular vectors with $\tau_{\alpha} \gtrsim $ 1 ns mostly have negative correlations $A_\alpha < 0$, manifesting anti-phase oscillations between the two output beams from \Lim cavities. In the elliptic microdisk lasers, however, the singular vectors of similar $\tau_{\alpha}$ do not exhibit any bias towards negative correlations, indicating the absence of alternate oscillations between output beams.

	
	We attribute the anti-correlated oscillations of the \Lim lasers to temporal beating of nearly degenerate modes. These modes interact nonlinearly through the gain medium, leading to their phase locking ~\cite{harayama2003asymmetric, fukushima2005unidirectional, choi2008alternate, shinohara2014anticorrelated}. When two lasing modes with slightly different frequencies and different mirror symmetry are phase-locked, their temporal beating causes anti-phased intensity oscillations of the output beams. The oscillation period is on the order of a few nanoseconds, which is much shorter than that of $\gtrsim$ 100 ns in quasi-stadium cavities \cite{choi2008alternate, shinohara2014anticorrelated}. To check whether anti-phase oscillations also occur with a longer period in \Lim cavities, we increase the time window of a streak image from 10~ns to 50~ns. The SVD analysis reveals alternate oscillations with period $\geq$ 10 ns (See Supplement 1). On even longer time scales, thermal effects become dominant, turning on or off lasing modes (Fig.~\ref{fig2}).    
	
	Compared to \Lim microdisk lasers, sub-nanosecond intensity fluctuations due to filament-like instabilities are more pronounced in elliptic cavities. One possible explanation is that WG modes in an elliptic cavity have a well-defined direction of propagation along the boundary. A small transverse wavevector (perpendicular to the local propagation direction) corresponds to a large transverse wavelength. Intensity variation on such length scales can induce lensing and self-focusing via carrier-induced index changes, which lead to filamentation and instability ~\cite{bittner2018suppressing}. In contrast, the chaotic ray dynamics in the \Lim cavity broadens the angular range of local propagation directions in a lasing mode, creating a more disordered intensity distribution. The decrease of transverse wavelength weakens the lensing effect, thus reducing filament-like lasing instabilities.   
	
	To summarize, we demonstrate the coexistence of multiple dynamic processes with distinct spatio-temporal scales in semiconductor microdisk lasers. The dominant process in a \Lim-shape cavity is alternate oscillations between two output beams. In an elliptic cavity, filament-like instabilities dominate the lasing dynamics. Theoretical studies are needed to understand better how microcavity geometry and ray dynamics influence the lasing dynamics and how different dynamical processes affect each other. An in-depth understanding will pave the way for using the microcavity shape to control the lasing dynamics.

\section*{Acknowledgments}
We acknowledge Ortwin Hess and Stefano Guazzotti for fruitful discussions. This work is supported by the US Naval Office of Research under Grant No. N00014-221-1-2026. S.\ B.\ acknowledges support for the Chair in Photonics from Minist\`ere de l'Enseignement Sup\'erieur, de la Recherche et de l'Innovation; R\'egion Grand-Est; D\'epartement Moselle; European Regional Development Fund (ERDF); Metz M\'etropole; GDI Simulation; CentraleSup\'elec; Fondation CentraleSup\'elec.

\bigskip
\renewcommand{\theequation}{S\arabic{equation}}
\renewcommand{\thefigure}{S\arabic{figure}}
\renewcommand{\thetable}{S\arabic{table}}
\setcounter{figure}{0}
\setcounter{equation}{0}
\setcounter{section}{0}

\section*{SUPPLEMENTARY MATERIAL}

\subsection*{Device fabrication and testing}

The microdisk lasers are fabricated on a commercial GaAs/AlGaAs quantum-well wafer (Q-Photonics QEWLD-808) by photolithography and inductively coupled plasma etching. The etch depth is about 3.5 $\mu$m \cite{bittner2018suppressing}. Top and bottom metal contacts are deposited for electric current injection. The top contact boundary is receded 6 $\mu$m from the disk edge to avoid it hanging down and blocking emission from the sidewall \cite{kim2022sensitive}. 

The semiconductor lasers are mounted on a copper mount. A tungsten needle (Quater Research H-20242) is placed on the top metal contact for electric current injection. The current is provided by a laser diode driver (DEI Scientific PCX-7401). To reduce heating, the electric current is injected with a pulse length of 2 $\mu$s at a repetition rate of less than 1 Hz. 

The emission from the edge of a semiconductor disk laser is collected by an objective lens (20$\times$, NA 0.4) and directed to an imaging spectrometer (Acton SP300i) with an intensified CCD camera (ICCD, Andor iStar DH312T-18U-73). To measure the time-resolved emission spectra, we time-gate the ICCD camera mounted on the output port of a spectrometer. 

Using a streak camera (Hamamatsu C5680) with a fast single sweep unit (M5676), we measure the time-resolved near-field intensity profiles. For \Lim-shaped cavities, two output beams are collected from one side of the disk. Edge emission from an elliptic cavity is strongest at the two vertices, leading to four output beams in two opposite directions. We measure two output beams from the elliptic microlasers by collecting the emission from one side of the disk.

\subsection*{Numerical simulation}

We use COMSOL Multiphysics \cite{comsol51} to calculate resonant modes in semiconductor microdisks. While the cavity shape is identical to the fabricated one, the cavity size is 10 times smaller to limit the computational load. The effective refractive index of the disk is 3.37. The polarization is transverse-electric (TE) with electric field parallel to the cavity plane, identical to that of edge-emission from a GaAs quantum-well laser. The mode shown in Fig.~1(b) of the main text has a Q-factor of 2.55$\times10^5$ and a wavelength of 798.71 nm. The whispering-gallery mode in Fig.~1(d) has a Q-factor of 3.76$\times10^{11}$ and a wavelength of 801.42 nm.

\subsection*{Thermal effect}
\label{thermal}

\begin{figure}[t]
	\centering\includegraphics[width=\linewidth]{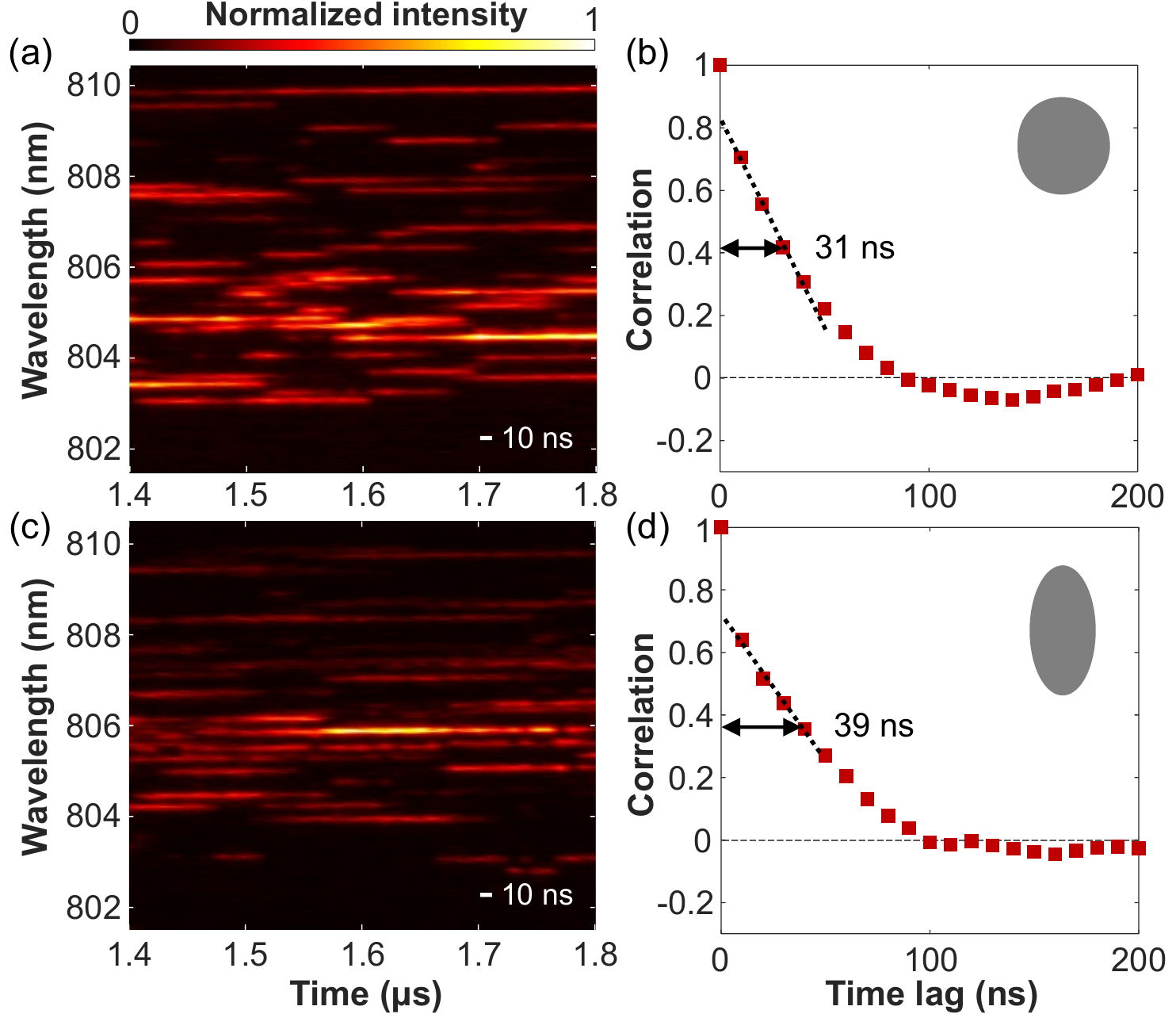}
	\caption{Temporal evolution of lasing spectrum due to heating. (a, c) Time-resolved spectrum of \Lim-shaped (a) and elliptic (c) microdisk lasers. The cavity dimensions are shown in Fig.~1, and the pump current is 500~mA. The temporal resolution determined by the gate time of the ICCD is 10 ns. (b, d) Spectral correlation function computed from the spectra in (a, c). The width of the correlation function gives the correlation time of 31 ns for \Lim (b) and 39 ns for elliptic (d) microcavity lasers.}
	\label{figS1}
\end{figure}

Figures 2(a) and (b) in the main text show the time-gated emission spectra taken with 50 ns long windows during the whole of 2 $\mu$s long pump pulse for \Lim and elliptic cavities. They reveal the temporal evolution of the emission spectrum due to laser heating. To investigate how rapidly the spectrum changes, we perform time-gated measurements with improve temporal resolution of 10 ns. The time-resolved lasing spectra at 500~mA (6.3 times the lasing threshold) are shown in Fig.~\ref{figS1} for a \Lim cavity (a) and an elliptic cavity (c). Over the interval of 400 ns, different lasing modes turn on and off as the emission spectrum shifts to a longer wavelength due to temperature increase.

\begin{figure*}[t]
	\centering\includegraphics[width=0.7\linewidth]{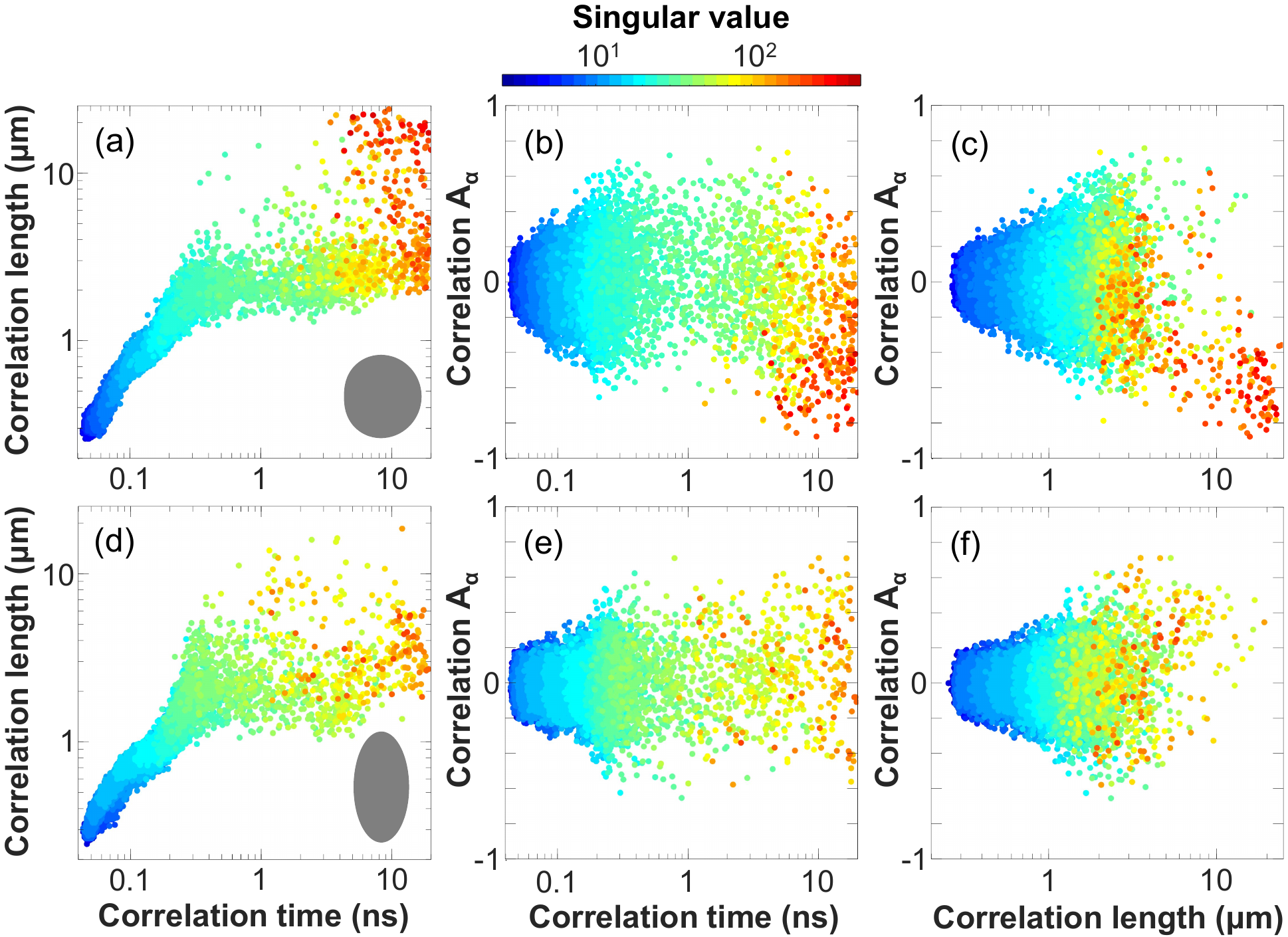}
	\caption{Spatio-temporal scales of intensity fluctuations in 50 ns long streak images of (a-c) \Lim-shaped microcavity lasers and (d-f) elliptic microcavity lasers. 
	(a, d) Spatial and temporal correlation widths of all singular vectors. 
	(b, e) Correlation $A_\alpha$ of temporal fluctuations of intensities of the two output beams vs. correlation time of individual singular vectors.
	(c, f) $A_\alpha$ vs. correlation length of individual singular vectors. The color represents the magnitude of the corresponding singular value. 
	For \Lim cavities, singular vectors with large singular values (in red color) concentrate around 10 ns and feature anti-correlated intensity fluctuations $A_\alpha < 0$, which are absent in elliptic cavities.}
	\label{figS2}
\end{figure*}

To characterize the time scale of lasing spectrum change due to heating, we calculate the spectral correlation function~\cite{bittner2018suppressing} defined by,
\begin{equation}
C(\tau) = \langle \delta I(\lambda, t) \delta I(\lambda, t+\tau) \rangle_{\lambda, t}
\label{eq:speccorr} \, ,
\end{equation}
where $\delta I(\lambda, t) = [I(\lambda, t)-\langle I(\lambda, t) \rangle_t] / \sigma_I(\lambda)$, and $\sigma_I(\lambda)= \sqrt{\langle [\delta I(\lambda, t)]^2 \rangle_{t}}$ . Figures~\ref{figS1}(b) and (d) show $C(\tau)$ for \Lim-shaped and elliptic microdisk lasers. The sharp drop around $\tau$ = 0 results from  photo-detection noise. It is followed by a gradual decay with $\tau$. To determine the correlation time, we extrapolate the slow decay to $\tau$ = 0 (denoted by a dotted line) and compute the half-width at half-maximum (HWHM) with respect to the extrapolated value at $\tau = 0$. The spectral correlation time is 31~ns for the \Lim-shaped microdisk laser [Fig.~\ref{figS1}(b)], and 39~ns for the elliptic microdisk laser [Fig.~\ref{figS1}(d)]. Please note that these time scales are considerably longer than the intrinsic time scales of the laser dynamics which are on the (sub-) nanosecond scale.

\subsection*{Lasing dynamics}

In Fig.~4(a), the correlation times of the singular vectors for lasing intensity fluctuations in \Lim-shaped cavities have an upper bound of 5 ns, set by the finite time range (10 ns) of the streak images. We performed additional measurements with a longer time range of 50 ns at the detriment of the temporal resolution which increases from 0.03 ns to 0.15 ns. We perform the singular value decomposition (SVD) analysis~\cite{kim2022sensitive} of streak images with a 50 ns range.  

Figure~\ref{figS2} shows spatial and temporal scales of singular vectors in \Lim-shaped [Figs.~\ref{figS2}(a-c)] and elliptic [Figs.~\ref{figS2}(d-f)] microcavity lasers. They are obtained from 68 streak images taken with two cavities of the same shape. For every singular vector, we compute the correlation time and the correlation length as well as the intensity correlation $A_{\alpha}$ between the two output beams [See Eq.~(1)]. 

The strongest singular vectors for \Lim-shaped cavities have a correlation time of around 10 ns [Fig.~\ref{figS2}(a)]. They are clustered in two groups: one with a correlation length longer than 10~$\mu$m, and the other with only a few $\mu$m. Figures~\ref{figS2}(b, c) reveal that the first group exhibits a negative correlation of intensity fluctuations between the two output beams, corresponding to anti-phase oscillations. The second group has weaker correlations (smaller $|A_\alpha|$) and an almost even distribution of $A_\alpha$ over positive and negative values. We attribute these singular vectors to the thermally-induced fluctuations (See Section~\ref{thermal}).

For elliptic microcavity lasers [Figs.~\ref{figS2}(d-f)], the dominant singular vectors are evenly distributed over positive ($A_\alpha>0$) and negative ($A_\alpha<0$) correlations. They may originate from lasing mode switching due to heating (Section~\ref{thermal}). Compared to Fig.~4(b), the singular vectors with sub-nanosecond correlation times are less strong because of the reduced temporal resolution of the streak images used here.

\bigskip\bigskip\bigskip\bigskip



\begin{thebibliography}{1}

\bibitem{cao2015dielectric} H.~Cao and J.~Wiersig, {\textit{Rev. Mod. Phys.}} \textbf{87}, 61 (2015).

\bibitem{schwefel2004progress} H.~Schwefel, H.~Tureci, A.~D. Stone, and R.~Chang, ``Progress in asymmetric resonant cavities: Using shape as a design parameter in dielectric microcavity lasers,'' in \textit{Optical Microcavities},  (World Scientific, 2004), pp. 415--495.

\bibitem{harayama2011two} T.~Harayama and S.~Shinohara, {\textit{Laser \& Photonics Rev.}} \textbf{5}, 247 (2011).

\bibitem{wiersig2008combining} J.~Wiersig and M.~Hentschel, {\textit{Phys. Rev. Lett.}} \textbf{100}, 033901 (2008).

\bibitem{wang2009deformed} Q.~J. Wang, C.~Yan, L.~Diehl, M.~Hentschel, J.~Wiersig, N.~Yu, C.~Pfl{\"u}gl,  M.~A. Belkin, T.~Edamura, M.~Yamanishi \emph{et~al.}, {\textit{New J. Phys.}} \textbf{11}, 125018 (2009).

\bibitem{shinohara2009ray} S.~Shinohara, M.~Hentschel, J.~Wiersig, T.~Sasaki, and T.~Harayama, {\textit{Phys. Rev. A}} \textbf{80}, 031801 (2009).

\bibitem{song2009chaotic} Q.~Song, W.~Fang, B.~Liu, S.-T. Ho, G.~S. Solomon, and H.~Cao, {\textit{Phys. Rev. A}} \textbf{80}, 041807 (2009).

\bibitem{yan2009directional} C.~Yan, Q.~J. Wang, L.~Diehl, M.~Hentschel, J.~Wiersig, N.~Yu, C.~Pfl{\"u}gl, F.~Capasso, M.~A. Belkin, T.~Edamura \emph{et~al.}, {\textit{Appl. Phys. Lett.}} \textbf{94}, 251101 (2009).

\bibitem{yi2009lasing} C.-H. Yi, M.-W. Kim, and C.-M. Kim, {\textit{Appl. Phys. Lett.}} \textbf{95}, 141107 (2009).

\bibitem{song2010directional} Q.~Song, L.~Ge, A.~Stone, H.~Cao, J.~Wiersig, J.-B. Shim, J.~Unterhinninghofen, W.~Fang, and G.~Solomon, {\textit{Phys. Rev. Lett.}} \textbf{105}, 103902 (2010).

\bibitem{albert2012directional} F.~Albert, C.~Hopfmann, A.~Ebersp{\"a}cher, F.~Arnold, M.~Emmerling, C.~Schneider, S.~H{\"o}fling, A.~Forchel, M.~Kamp, J.~Wiersig \emph{et~al.}, {\textit{Appl. Phys. Lett.}} \textbf{101}, 021116 (2012).

\bibitem{redding2012local} B.~Redding, L.~Ge, Q.~Song, J.~Wiersig, G.~S. Solomon, and H.~Cao, {\textit{Phys. Rev. Lett.}} \textbf{108}, 253902 (2012).

\bibitem{kreismann2017three} J.~Kreismann, S.~Sinzinger, and M.~Hentschel, {\textit{Phys. Rev. A}} \textbf{95}, 011801 (2017).

\bibitem{shim2011whispering} J.-B. Shim, J.~Wiersig, and H.~Cao, {\textit{Phys. Rev. E}} \textbf{84}, 035202 (2011).

\bibitem{shim2013adiabatic} J.-B. Shim, A.~Ebersp{\"a}cher, and J.~Wiersig, {\textit{New J. Phys.}} \textbf{15}, 113058 (2013).

\bibitem{kraft2014perturbative} M.~Kraft and J.~Wiersig, {\textit{Phys. Rev. A}} \textbf{89}, 023819 (2014).

\bibitem{choi2008alternate} M.~Choi, T.~Fukushima, and T.~Harayama, {\textit{Phys. Rev. A}} \textbf{77}, 063814 (2008).

\bibitem{shinohara2014anticorrelated} S.~Shinohara, T.~Fukushima, S.~Sunada, T.~Harayama, K.~Arai, and K.~Yoshimura, {\textit{Opt. Rev.}} \textbf{21}, 113 (2014).

\bibitem{bittner2018suppressing} S.~Bittner, S.~Guazzotti, Y.~Zeng, X.~Hu, H.~Y{\i}lmaz, K.~Kim, S.~S. Oh, Q.~J. Wang, O.~Hess, and H.~Cao, {\textit{Science}} \textbf{361}, 1225 (2018).

\bibitem{kim2021massively} K.~Kim, S.~Bittner, Y.~Zeng, S.~Guazzotti, O.~Hess, Q.~J. Wang, and H.~Cao, {\textit{Science}} \textbf{371}, 948 (2021).

\bibitem{kim2022sensitive} K.~Kim, S.~Bittner, Y.~Jin, Y.~Zeng, S.~Guazzotti, O.~Hess, Q.~J. Wang, and H.~Cao, {\textit{APL Photonics}} \textbf{7}, 056106 (2022).

\bibitem{ma2022chaotic} C.-G. Ma, J.-L. Xiao, Z.-X. Xiao, Y.-D. Yang, and Y.-Z. Huang, {\textit{Light. Sci. \& Appl.}} \textbf{11}, 1 (2022).

\bibitem{nockel1996directional} J.~U. N{\"o}ckel, A.~D. Stone, G.~Chen, H.~L. Grossman, and R.~K. Chang, {\textit{Opt. Lett.}} \textbf{21}, 1609 (1996).

\bibitem{kim2004highly} S.-K. Kim, S.-H. Kim, G.-H. Kim, H.-G. Park, D.-J. Shin, and Y.-H. Lee, {\textit{Appl. Phys. Lett.}} \textbf{84}, 861 (2004).

\bibitem{schwefel2004dramatic} H.~G. Schwefel, N.~B. Rex, H.~E. Tureci, R.~K. Chang, A.~D. Stone, T.~Ben-Messaoud, and J.~Zyss, {\textit{J. Opt. Soc. Am. B}} \textbf{21}, 923 (2004).

\bibitem{hess1994spatio} O.~Hess, {\textit{Chaos, Solitons \& Fractals}} \textbf{4}, 1597 (1994).

\bibitem{fischer1996complex} I.~Fischer, O.~Hess, W.~Els{\"a}{\ss}er, and E.~G{\"o}bel, {\textit{Europhys. Lett.}} \textbf{35}, 579 (1996).

\bibitem{marciante1998spatio} J.~R. Marciante and G.~P. Agrawal, {\textit{IEEE Photonics Technol. Lett.}} \textbf{10}, 54 (1998).

\bibitem{harayama2003asymmetric} T.~Harayama, T.~Fukushima, S.~Sunada, and K.~S. Ikeda, {\textit{Phys. Rev. Lett.}} \textbf{91}, 073903 (2003).

\bibitem{fukushima2005unidirectional} T.~Fukushima, T.~Tanaka, and T.~Harayama, {\textit{Appl. Phys. Lett.}} \textbf{86}, 171103 (2005).

\bibitem{comsol51} \textit{COMSOL Multiphysics Reference Manual, version 5.1,} www.comsol.com.

\end{thebibliography}

\end{document}